\documentclass[aps, pra,twocolumn,showpacs,superscriptaddress]{revtex4}
\usepackage{amsmath}
\usepackage{graphicx}
\usepackage{dcolumn}
\usepackage{bm}
\begin{document}
\title{Bose-Einstein condensate in gases of rare atomic species} 
\author{Seiji Sugawa}
\email{ssugawa@scphys.kyoto-u.ac.jp}
\affiliation{Department of Physics, Graduate School of Science, Kyoto University 606-8502 Japan}
\author{Rekishu Yamazaki}
\affiliation{Department of Physics, Graduate School of Science, Kyoto University 606-8502 Japan}
\affiliation{CREST, JST, Chiyoda-ku, Tokyo 102-0075 Japan}
\author{Shintaro Taie}
\affiliation{Department of Physics, Graduate School of Science, Kyoto University 606-8502 Japan}
\author{Yoshiro Takahashi}
\affiliation{Department of Physics, Graduate School of Science, Kyoto University 606-8502 Japan}
\affiliation{CREST, JST, Chiyoda-ku, Tokyo 102-0075 Japan}
\date{\today}
\begin{abstract}
We report on the successful extension of production of Bose-Einstein Condensate (BEC) to rare species.
Despite its low natural abundance of 0.13\%, $^{168}$Yb is directly evaporatively cooled down to BEC.
Our successful demonstration encourages attempts to obtain quantum gases of radioactive atoms,
which extends the possibility of quantum many-body physics and precision measurement.
Moreover, a stable binary mixture of $^{168}$Yb BEC and $^{174}$Yb BEC is successfully formed.
\end{abstract}
\maketitle
\parskip=0pt
Successful realization of Bose-Einstein condensates (BEC) and Fermi degenerate gases in atomic vapors has been providing intriguing possibilities in many research fields
such as the study of quantum many-body system, precision measurement, and quantum computation(see Ref.\cite{BlochReview} for review).
So far atomic species with relatively high natural abundance have been Bose-Einstein condensed,
such as in alkali-atoms$^{87,85}$Rb\cite{87Rb, 85Rb}, $^{39,41}$K\cite{39K, 41K}, $^{7}$Li\cite{7Li}, $^{23}$Na\cite{23Na} and $^{133}$Cs\cite{133Cs},  
and two-electron atoms such as $^{4}$He\cite{Robert2001, Santos2001}, $^{174, 170, 176}$Yb\cite{Takasu174Yb, Fukuhara170Yb, FukuharaMix}, 
$^{40}$Ca\cite{Kraft40Ca}, and $^{84,86,88}$Sr\cite{Stellmer84Sr, Escobar84Sr, 86Sr, 88Sr},
and also $^{1}$H\cite{HydBEC} and $^{52}$Cr\cite{52Cr}.
Extending atomic species beyond these relatively abundant atomic species is an important step for a future investigation.
First, extension to rare species benefits a richer variety of quantum degenerate gases and the mixed gases in the study of quantum many-body system.
In addition, radioactive atoms such as francium\cite{Fr2, Fr} and radium\cite{Ra2007} are good candidates
for tests of the fundamental symmetries by searching for the electric dipole moment and precision measurement of atomic parity non-conservation.
In such application, a Mott-insulator state in an optical lattice, realized by making use of quantum gases, offers important advantage of high atomic density with avoided collision by localizing each atom to each lattice site.

Obviously, the limited abundance of atoms is a great obstacle to reach quantum degeneracy\cite{enrich}.
However, in principle, there is no fundamental difficulty to achieve quantum degeneracy of rare species by evaporative cooling if the ratio of elastic to inelastic scattering cross section is high.
In this Rapid Communication, we report on the successful production of BEC of rare species of $^{168}$Yb,
which has the lowest natural abundance of only 0.13\% among the seven stable isotopes of Yb without the use of enriched source.
The abundance of 0.13\% is also the lowest among the already produced BEC\cite{enrich}.
The atom clouds can be directly evaporatively cooled down to BEC owing to a favorable s-wave scattering length of $^{168}$Yb and a high vacuum of our system.
The lowest natural abundant species so far was $^{84}$Sr, which has a natural abundance of 0.56\%.
The difference from $^{84}$Sr BEC is the small initially-captured atom number.
While $10^8$ atoms were magnetically trapped in $^{84}$Sr, only $2 \times 10^6$ atoms are captured in a magneto-optical trap (MOT) for $^{168}$Yb.
This value is, for example, not too far from the atom number of radioactive francium MOT\cite{largeMOT}.
Therefore, our successful demonstration encourages attempts to obtain quantum gases of radioactive atoms\cite{radioBEC}.
A straightforward implication of our result is the promising possibility of realizing quantum degenerate gases of radioactive Yb isotopes for precision measurement\cite{radioYb}. 

In addition, we report on the attainment of dual-species condensates of $^{168}$Yb and $^{174}$Yb.
This is the first stable binary condensate for a two-electron system.
Since two species BEC has been reported only for $^{85}$Rb and $^{87}$Rb\cite{Papp8587}, $^{41}$K and $^{87}$Rb\cite{ModugnoRbK}, 
and quite recently $^{87}$Rb and $^{133}$Cs mixtures\cite{LercherRbCs}, 
the demonstrated dual condensate of $^{168}$Yb and $^{174}$Yb in fact extends the variety of the quantum degenerate mixture, 
where quantum turbulences\cite{turbulence} and topological excitations\cite{Eto2011} can be studied.

In order to produce $^{168}$Yb BEC, we have revised our experimental system from the previously reported ones
\cite{Takasu174Yb, Fukuhara173Yb, Fukuhara170Yb, FukuharaMix}.
The major improvement is the high vacuum of 1$\sim$3$\times$10$^{-11}$ Torr compatible with the highly collimated atomic beam.
The experimental procedure to produce $^{168}$Yb BEC follows our previous scheme\cite{Takasu174Yb, Fukuhara173Yb, Fukuhara170Yb, FukuharaMix,Taie2010}.
First, an $^{168}$Yb atomic beam from an oven containing Yb metal with natural abundance  is decelerated by a Zeeman slower 
operating on the ${}^1{\rm S}_0 \leftrightarrow {}^1{\mathrm P}_1$ transition with a laser power of 250 mW.
The decelerated atoms are collected by a MOT operating on the narrow-line inter-combination transition of ${}^1{\rm S}_0 \leftrightarrow {}^{3}{\rm P}_1$. 
Our high vacuum system enables us to typically collect 2$\times$10$^6$ $^{168}$Yb atoms at a temperature of 20 $\mu$K during the 100 s loading.
This should be compared with the case of the most abundant isotope of $^{174}$Yb where it takes less than 0.5 s to collect 2 $\times$ 10$^6$ atoms.
The loading time of about 100 s would be limited by a very little branching from the  ${}^3{\rm P}_1$ state to the ${}^3{\rm P}_0$ state through the magnetic dipole transition
at a rate of $\sim$ 1/161 s$^{-1}$\cite{branching} and also by background gas collisions.

Then we transfer the atoms from the MOT to a Far-Off Resonant Trap (FORT) operating at 532 nm to perform efficient evaporative cooling.
Efficient transfer is performed by compressing the magnetic field gradient of the MOT and subsequently reducing the MOT intensity and the detuning of MOT laser beams to cool the atoms.
Typically  1$\times$10$^6$ $^{168}$Yb atoms are successfully trapped into the FORT.

The evaporative cooling is performed in a crossed FORT, where horizontally and vertically propagating FORT beams intersect at their foci.
The horizontally propagating FORT beam is elliptically shaped with a beam waist of 30 $\mu$m along the horizontal direction and  14 $\mu$m along the vertical direction.
The initial trap depth is approximately 750 $\mu$K at the horizontal FORT laser power of 14 W.
The first stage of evaporative cooling is performed by lowing the laser intensity of the horizontal FORT beam,
during which the atoms are transfered to the crossed FORT region.
The shape of the vertically propagating FORT beam is also elliptical with a beam waist of  127 $\mu$m along the axial direction of the horizontal FORT beam and 40 $\mu$m along the radial.
This results in a good overlap of the two FORT beams with an increased trap volume.
The potential depth by the vertical FORT beam is 40 $\mu$K.
At 40 $\mu$K potential depth of the crossed FORT, 5 $\times$ 10$^5$ atoms at 6 $\mu$K are trapped in the crossed region.

The atoms in the crossed FORT region are further cooled down toward the quantum degenerate regime at the final stage of the evaporative cooling.
The s-wave scattering length of $^{168}$Yb is determined to be $a_{168} = 13.33(18)$ nm from two-photon photoassociation spectroscopy\cite{Kitagawa2PA}.
The evaporative cooling should work well at this scattering length.
To suppress unfavorable inelastic three-body loss, we not only reduce the horizontal FORT beam intensity, but also the vertical FORT beam intensity during the evaporation to reduce the central peak density. 
The peak density above the BEC phase transition is typically 2$\sim$4$\times$10$^{13}$ cm$^{-3}$. 

		\begin{center}
		\begin{figure}[thb]
		\includegraphics[width=8cm]{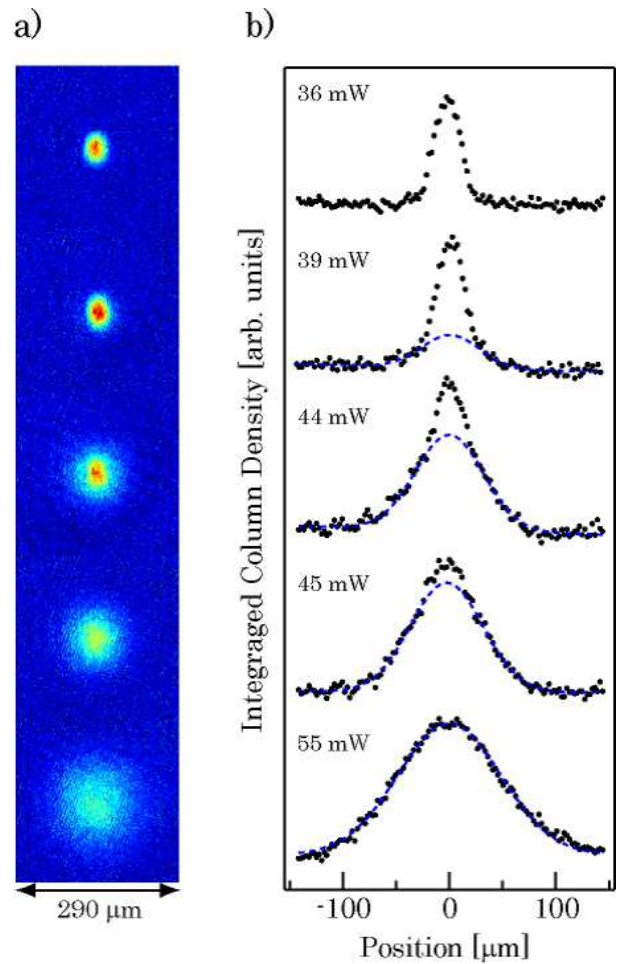}
		\caption{(Color online) Phase transition to BEC during the evaporative cooling.
			(a) Absorption images at different evaporation stage are shown with TOF of 14 ms.
			Each image is 290 $\mu$m per side.
			(b) The corresponding integrated column densities along the horizontal direction are shown with laser powers of horizontal FORT beam.
			The BEC transition is observed as an appearance of bimodal distribution (middle). 
			At the end of the evaporation, a almost pure condensate of 1.0 $\times$ 10$^4$ is produced (top).
			The absorption images are taken from an angle of 45 degree against the horizontal FORT axis in the horizontal plane.
			}
		\label{bimodal}
		\end{figure}
		\end{center}

The phase transition from a thermal gas to a BEC is observed as the appearance of bimodal distribution after Time-of-Flight (TOF).
Figure 1 shows absorption image taken after TOF of 14 ms for different evaporation stage near BEC phase transition.
At a higher temperature, the density distribution shows an isotropic Gaussian distribution, manifesting that the velocity of the gas has a Maxwell-Boltzmann distribution.
However, after further evaporative cooling, we observe the deviation from the Gaussian distribution and an additional low velocity component, representing a BEC component. 
At this stage, the peak density of the thermal gas is 3.5 $\times$ 10$^{13}$ cm$^{-3}$.
The trap frequencies are 2 $\pi$ $\times$ (240, 31 165) Hz.
The transition temperature is evaluated to be $T_c$ = 135 nK.
This value is slightly lower than the transition temperature for non-interacting BEC, 
$T_{\rm c}= \hbar \omega / {\rm k_B} \times (N/\zeta(3))^{1/3} \sim$ 150 nK, 
where $\omega$ is the geometric mean of trap frequency, $\zeta$ is the  Riemann zeta function, and $N$ is the atom number. 
This slight deviation is expected from the interaction and finite size effects\cite{Giorgini1996}.

After further evaporation by lowing the trap depth, the condensate fraction increases.
After evaporative cooling for 9.3 s, we finally attain  an almost pure condensate up to 1.0 $\times$ 10$^4$ atoms.
The trap frequencies are 2 $\pi$ $\times$ (154, 25, 141) Hz.
The peak density is 1.0 $\times $10$^{14}$ cm$^{-3}$.
The validity of Thomas-Fermi approximation is confirmed by the relation 
$N a_{168}/a_{\rm ho}=1.4 \times 10^2 \gg 1$, where $a_{\rm ho}=\sqrt{\hbar/(m \omega)}$ is a harmonic oscillation length and $m$ is the atomic mass.
Figure 2 shows radii of the expanding BEC during the TOF.
Anisotropic expansion of condensate can be clearly seen.
The solid lines in Fig.1 are the fits using the Castin-Dum model with the chemical potential as the only free parameter. 
The corresponding chemical potential is 64 nK.
It is noted that, with this achievement, all the isotopes of Yb are now cooled down to quantum degenerate regime except for the bosonic isotope of $^{172}$Yb,
where a large number of condensate cannot grow due to its large negative scattering length\cite{Kitagawa2PA}.

At the final trap depth, a lifetime of BEC is evaluated to be about 3 s, which is mainly attributed to the three-body loss of atoms.
From this, we can determine the upper limit of three-body loss rate of $^{168}$Yb.
We fit the remaining atom number as a function of holding time in the trap by a rate equation assuming only for three-body loss.
The maximum three-body coefficient $K_3$ is evaluated to be  8.6(1.5) $\times$ 10$^{-28}$ cm$^{6}$/s. 
This is more than one order of magnitude larger than the value reported for $^{174}$Yb\cite{Takasu174Yb, FukuharaMix}.
However it is still smaller than the expected maximal rate of $K_{3,max}=210 \hbar {a_{168}}^4/m=2.3 \times 10^{-27}$cm$^{6}$/s\cite{BedaqueK3,FedichevK3}.

		\begin{center}
		\begin{figure}[tbh]
		\includegraphics[width=6cm]{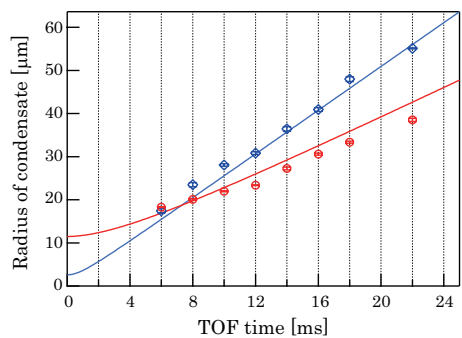}
			\caption{(Color online) Expansion of condensate after release from the trap.
		The radii of the condensate show asymmetry between the horizontal (circle) and vertical (square) direction.
		The solid lines are the fits using Castin-Dum model. The error bars are standard error.}
		\end{figure}
		\end{center}

Our successful demonstration of BEC of rare species extends the possibility to create various quantum degenerate mixtures.
This is especially important for Yb quantum isotope mixture, since a stable BEC mixture has not been yet reported.
Previously, we reported on the production of Bose-Bose mixture of $^{174}$Yb-$^{176}$Yb below T$_c$\cite{FukuharaMix}.
However, due to the negative scattering length of $^{176}$Yb, a condensate above the critical atom number (typically a few hundred) gets unstable.
Among the possible quantum degenerate Bose-Bose mixtures of Yb isotope, $^{168}$Yb-$^{174}$Yb and $^{168}$Yb-$^{170}$Yb
are the only candidate to realize a stable BEC mixture.
Here, we report on the production of stable $^{168}$Yb-$^{174}$Yb Bose-Bose mixture.

The procedure to produce this mixture is basically the same as that describe above.
To collect two isotopes in the MOT, we apply two-color MOT beam.
Due to the large difference of the atomic beam flux, originates from the large difference in the natural abundance of about a factor of 230,
we load $^{168}$Yb for 100 s 
and then load $^{174}$Yb for 0.5 s by subsequently turning on the Zeeman slower beam.
Since the interspecies scattering length between the two species is as small as $0.13(18)$nm,   
 both species are almost independently evaporatively cooled to quantum  degenerate regime.
Figure 3 shows the absorption images of the dual condensate.
The number of atoms in each condensate is 0.9$\times$10$^4$.
The advantage of working with Yb isotope mixture is the fact that trap parameters such as trap depth, trap frequency, and gravitational sag, are almost the same for the different isotopes.
This enables us to focus on the role of interaction.
Such feature is already seen in the behavior of the expanding condensate in Fig. 3, 
where $^{168}$Yb is expanding much faster than $^{174}$Yb.
As the interspecies interaction is small, the difference in the size can be explained by the mean field energy of each condensate, 
which is converted into kinetic energy during the expansion.
The expansion energy is conserved to 2$\mu$/7, which is proportional to $(N a_{M})^{2/5}$ in the Thomas-Fermi approximation,
where $N$ is the number of condensate, $\mu$ is the chemical potential and $a_{M}$ is the scattering length with mass number $M$. 
The expansion velocity normalized by $N^{1/5}$ of $^{168}$Yb condensate is evaluated to be 1.20(4) times faster than $^{174}$Yb from the TOF image of each condensate.
This is in good agreement with the ratio of scattering length ($a_{168}$/$a_{174}$)$^{1/5}$=1.19. 
The ratio of each number of atoms can be easily tuned by changing the loading time to the MOT.
We can so far prepare an almost pure Bose-Bose mixture with $^{174}$Yb up to 7 $\times$ 10$^4$.
Although the s-wave interspecies scattering length is small, optical Feshbach resonance will enable tuning of interspecies interaction,  
which is demonstrated for BEC of $^{174}$Yb\cite{Yamazaki2010} and thermal gases of $^{172}$Yb and $^{176}$Yb\cite{EnomotoOFR}.
Study of quantum quench dynamics and non-equilibrium dynamics\cite{quench} in binary condensate system will also be accessible.

		\begin{center}
		\begin{figure}[bht]
		\includegraphics[width=7cm]{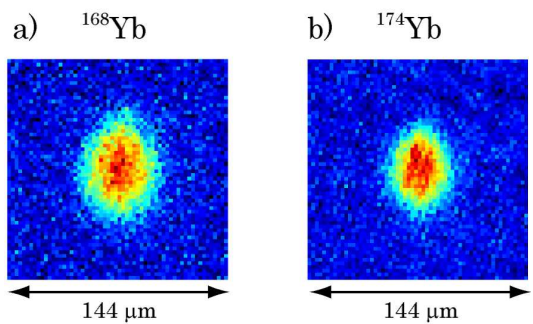}
		\caption{(Color online) A stable Bose-Bose mixture of $^{168}$Yb-$^{174}$Yb.
			Absorption images of 
			(a) almost pure $^{168}$Yb BEC of 9$\times$10$^3$ and 
			(b) almost pure $^{174}$Yb BEC  of 9$\times$10$^3$. 
			Both images are taken  after TOF time of 18 ms in a separate experimental run. 
			The larger cloud size of $^{168}$Yb compared with $^{174}$Yb manifests larger mean field energy of the condensate of $^{168}$Yb compared with $^{174}$Yb. 
			}
		\label{BBMix}
		\end{figure}
		\end{center}

In conclusion, we have produced a BEC of rare species of $^{168}$Yb containing 1.0 $\times$ 10$^4$ atoms. 
With this achievement all the possible isotopes of Yb have been cooled to quantum degeneracy. 
Despite its low natural abundance of 0.13\%, $^{168}$Yb is efficiently collected and cooled down to BEC.
Our demonstration encourages researchers that a radioactive isotope is a realistic candidate of BEC.
Moreover, $^{168}$Yb BEC is produced together with $^{174}$Yb BEC as a stable binary BEC.
This mixture, which has very small interspecies scattering length, will be a prime candidate 
for studying a binary BEC, including the possible use of optical Feshbach resonances. 

This work is supported by the Grant-in-Aid for Scientific Research of JSPS (No. 18204035, 21102005C01 (Quantum Cybernetics)), 
GCOE Program "The Next Generation of Physics, Spun from Universality and Emergence" from MEXT of Japan,
and World- Leading Innovative R\&D on Science and Technology (FIRST).
S.S and S.T acknowledge support from JSPS.


\end{document}